\documentstyle[12pt]{article}

\begin{document}

\begin{flushright}
IMSc/2005/03/04 \\
arXiv: hep-th/0503058
\end{flushright} 

\vspace{2mm}

\vspace{2ex}

\begin{center}
{\large \bf Multiparameter Brane Solutions  } \\ 

\vspace{2ex}

{\large \bf by Boosts, S and T dualities } \\

\vspace{8ex}

{\large  S. Kalyana Rama}

\vspace{3ex}

Institute of Mathematical Sciences, C. I. T. Campus, 

Taramani, CHENNAI 600 113, India. 

\vspace{1ex}

email: krama@imsc.res.in \\ 

\end{center}

\vspace{6ex}

\centerline{ABSTRACT}
\begin{quote} 

We show that the multiparameter (intersecting) brane solutions
of string/M theories given in the literature can all be obtained
by a suitable combination of boosts in eleven dimension, S and T
dualities. We also describe a duality property of the $D$
dimensional multiparameter solutions describing branes smeared
in compact directions. This duality is analogous to the T
duality of the string theory but is valid for any value of $D$.

\end{quote}

\newpage

\vspace{2ex}

\begin{center}
{\large \bf 1. Introduction}
\end{center}

\vspace{2ex}

Recently Zhou and Zhu obtained multiparameter brane solutions
\cite{zz}. For particular values of the parameters, these
solutions reduce to the well known brane solutions \cite{solns}
of string/M theories. These multiparameter solutions were
interpreted in \cite{bmo} as describing non BPS brane antibrane
configurations and were used to study the tachyon condensation
and the related dynamics. Since then, they have been generalised
and studied extensively \cite{kam, uses, glc, lr, mo}.

In all these cases, the multiparameter solutions have been
obtained by solving the relevent equations of motion directly
with varying levels of generality. The most general of such
solutions have been obtained by Miao and Ohta (MO) in \cite{mo}
which also includes intersecting brane configurations. In this
paper we show that all of these multiparameter solutions
describing intersecting brane configurations in string/M
theories can be obtained by a suitable combination of boosts in
eleven dimension, S and T dualities \cite{rules, boost, bho}.

Assuming an ansatz similar to that in \cite{bk} we first obtain
a generalisation of the solution in \cite{bk}, now including a
requisite number of compact directions also. Taking the
spacetime to be eleven dimensional with atleast one compact
direction, we boost this solution and reduce to ten dimensions
obtaining thereby a ten dimensional charged solution. Applying
further a suitable combination of S and T dualities, and
lift-boost-reduce procedures to generate more brane charges,
will then yield a variety of intersecting brane configurations
of string/M theories. The calculations are all straightforward
and we present a few brane solutions explicitly by way of
illustration. We also point out that starting from extremal
(intersecting) brane configurations, the corresponding
multiparameter solutions can be written down following a few
simple rules, similar to those given in \cite{rules}.

We show that the $D$ dimensional multiparameter solutions
describing branes smeared in compact directions have a duality
property when the higher form field strength corresponding to
the brane couples to the scalar field in a specific way. This
duality is analogous to the T duality of the string theory but
is valid for any value of $D$. For $D = 10$, and for $D < 10$
also, this is same as T duality of the string theory. For $D >
10$, this duality can be used as a solution generating technique
but otherwise its significance, if any, is not clear.

Although the solutions are obtained here assuming an ansatz as
in \cite{bk}, they turn out to be completely equivalent to the
corresponding ones in \cite{mo}, and hence also to those in
\cite{zz, bmo, kam, glc, lr}, which were obtained by solving the
relevent equations of motion directly. This is shown by writing
our solutions in isotropic coordinates, and then comparing with
those given in \cite{zz, bmo, kam, mo} after some elaborate
algebraic manipulations. \footnote{Along the way, it can be seen
that of the two parameters, $\mu$ and $\nu$, in the solutions of
\cite{mo} only one combination is physically relevent; the other
one amounts to a diffeomorphism.} It then follows that the
multiparameter brane solutions of string/M theories given in
\cite{zz, bmo, kam, uses, glc, lr, mo} can all be obtained by a
suitable combination of boosts in eleven dimensions, S and T
dualities.

This paper is organised as follows. In section {\bf 2} we
present the $(D + 1)$ dimensional uncharged multiparameter
solutions. In section {\bf 3} we use them to obtain $D$
dimensional uncharged solutions, and also the charged ones by a
boost in $(D + 1)$ dimensions. In section {\bf 4} we set $D =
10$ and write down explicitly a few multiparameter
(intersecting) brane solutions of string/M theories explicitly.
We then present a few simple rules for obtaining such solutions
starting from the corresponding extremal ones. In section {\bf
5} we describe a duality property of the $D$ dimensional
multiparameter solutions describing branes smeared in compact
directions. In an Appendix we show the equivalence of our
solutions to the corresponding ones in the literature. In
section {\bf 6} we conclude by mentioning a few issues for
further studies.

\vspace{2ex}

\begin{center}
{\large \bf 2. $(D + 1)$ dimensional solutions}
\end{center}

\vspace{2ex}

Consider the Einstein action 
\begin{equation}\label{shat}
\hat{S} = \frac{1}{2 \kappa^2 L} 
\int d^{D + 1}X \sqrt{- \hat{g}} \; \hat{{\cal R}} 
\end{equation}
in a $D + 1$ dimensional spacetime and its dimensional
reduction to $D$ dimensions along a compact spatial direction of
size $L$. Let the $D + 1$ dimensional line element be given by
\begin{equation}\label{dhattod}
d \hat{S}^2 = e^{- \frac{2 b \phi}{D - 2}} \; g_{\mu \nu} \; 
d X^\mu d X^\nu + e^{2 b \phi} \; (d z + A_\mu d X^\mu)^2 
\end{equation}
where $\mu, \nu = 0, 1, \cdots, (D - 1)$ and the fields $(g_{\mu
\nu}, \phi, A_\mu)$ are all independent of the coordinate $z$
along the compact direction. For $b = \sqrt{\frac{D - 2}{2 (D -
1)}}$ the action $\hat{S}$ in (\ref{shat}) reduces to the $D$
dimensional `Einstein frame' action $S$ for a scalar $\phi$ and
a $2-$form field strength $F_{2} = \partial_\mu A_\nu -
\partial_\nu A_\mu$ given by
\begin{equation}\label{s}
S = \frac{1}{2 \kappa^2} \int d^DX \sqrt{- g} 
\left( {\cal R} - \frac{1}{2} (\partial \phi)^2 
- \frac{e^{\lambda \phi} \; F_{2}^2 }{4} \right ) 
\end{equation}
where $\lambda = \frac{2 (D - 1) b}{D - 2} 
= \sqrt{\frac{2 (D - 1)}{D - 2}}$. 

The above action is a particular case of the more general $D$
dimensional Einstein frame action for $g_{\mu \nu}, \phi$, and a
$(p_1 + 1)-$form gauge field $A_{(p_1 + 1)}$ given by
\begin{equation}\label{sp}
S_{p_1} = \frac{1}{2 \kappa^2} \int d^DX \sqrt{- g} \left( 
{\cal R} - \frac{1}{2} (\partial \phi)^2 - \frac{ 
e^{\lambda_{p_1} \phi} \; F_{p_1 + 2}^2}{2 (p_1 + 2) !} 
\right ) 
\end{equation}
where $\lambda_{p_1}$ is an arbitrary constant which may depend
on $p_1$ and the $(p_1 + 2)-$form field strength $F_{p_1 + 2} =
d A_{(p_1 + 1)}$. For $p_1 = 0$ and the specific value of
$\lambda_0 = \lambda$, the action (\ref{sp}) reduces to action
(\ref{s}) and can therefore be obtained by dimensional reduction
of a $(D + 1)$ dimensional action (\ref{shat}).

In the following, we will consider electric type $p_1-$brane
solutions in the string/M theories. (Magnetic ones can be easily
obtained from these solutions.) Then $D =10$ in the actions
(\ref{shat}) and (\ref{s}). In the action (\ref{sp}), $D = 11$
$\lambda_p = 0$, and the field $\phi$ is absent for M theory
branes; whereas $D = 10$ for string theory with $\lambda_p = -
1, + 1$ for fundamental strings and the 5-branes respectively in
Neveu-Schwarz (NS) sector, and $\lambda_p = \frac{3 - p}{2}$ for
branes in the Ramond sector.

Now, consider $D + 1$ dimensional spacetime with $p + 1$ compact
spatial directions. Consider the time independent case with the
line element $d \hat{S}$ given by
\begin{equation}\label{ansatzhat}
d \hat{S}^2 = - e^{2 \alpha_0} d t^2 
+ \sum_{i = 1}^{p + 1} e^{2 \alpha_i} d X^{i 2} + 
e^{2 \gamma} d r^2 + e^{2 \omega} d \Omega_{n + 1}^2 
\end{equation}
where $n = D - 3 - p$, $r$ is the radial coordinate in the $(n +
2)$ dimensional non compact tranvserse space, $d \Omega_{n + 1}$
is the standard line element on an $(n + 1)$ dimensional unit
sphere, and the functions $(\alpha_i, \gamma, \omega)$ with $i =
0, 1, \cdots, p + 1$ depend on $r$ only. The independent non
zero components of the Riemann tensor $\hat{{\cal R}}^K_{\; \; L
M N}$ for the above ansatz are 
\begin{eqnarray}
\hat{{\cal R}}^a_{\; \; b c d} & = & (\delta^a_c 
\sigma_{b d} - \delta^a_d \sigma_{b c}) \; \omega_r^2 \; 
e^{2 (\omega - \gamma)} + \rho^a_{\; \; b c d} \nonumber \\ 
\hat{{\cal R}}^a_{\; \; i b j} & = & 
\delta^a_b \delta_{i j} \;  (\alpha_i)_r \omega_r \; 
e^{2 (\alpha_i - \gamma)} \nonumber \\  
\hat{{\cal R}}^a_{\; \; r b r} & = & \delta^a_b \; 
(\omega_{r r} + \omega_r^2 - \omega_r \gamma_r) \nonumber \\
\hat{{\cal R}}^i_{\; \; r j r} & = & \delta^i_j \; 
((\alpha_i)_{r r} + (\alpha_i)_r^2 - (\alpha_i)_r 
\gamma_r) \nonumber \\
\hat{{\cal R}}^i_{\; \; j k l} & = & 
(\delta^i_k \delta{j l} - \delta^i_l \delta{j k}) \; 
(\alpha_i)_r (\alpha_j)_r \; e^{2 (\alpha_j - \gamma)} 
\label{riemann}
\end{eqnarray}
where $\sigma_{a b}$ with $a, b = 1, 2, \cdots, n + 1$ is the
metric on the $(n + 1)$ dimensional unit sphere, $\rho^a_{\; \;
b c d} = (\delta^a_d \sigma_{b c} - \delta^a_c \sigma_{b d})$
its Riemann tensor, and $(\;)_r \equiv \frac{d}{d r} (\;)$. The
Ricci tensor $\hat{{\cal R}}_{M N}$ can be obtained from the
above expressions.

Action (\ref{shat}) gives the equations of motion $\hat{{\cal
R}}_{M N} = 0$. Using (\ref{riemann}), one can then obtain the
equations of motion for $(\alpha_i, \gamma, \omega)$. These
equations can be solved easily by assuming the ansatz 
\begin{equation}\label{ansatz}
e^{\alpha_i} = Z^{\hat{a}_i} \; , \; \; \; 
e^\gamma = Z^{\hat{b}}  \; , \; \; \; 
e^\omega = r \; Z^{\hat{c}}  \; , \; \; \; 
Z \equiv 1 - \frac{r_0^n}{r^n} 
\end{equation}
where $r_0$ and $(\hat{a}_i, \hat{b}, \hat{c})$ are constant
parameters. Equations $\hat{{\cal R}}_{a b} = 0$ imply that
$\hat{b} = \hat{c} - \frac{1}{2}$. Thus, the line element in
(\ref{ansatzhat}) becomes
\begin{equation}\label{d+1}
d \hat{S}^2 = - Z^{2 \hat{a}_0} d t^2 + \sum_{i = 1}^{p + 1}
Z^{2 \hat{a}_i} d X^{i 2} + Z^{2 \hat{c}} d s_{n + 2}^2
\end{equation}
where we have defined 
\begin{equation}\label{dsn+2}
d s_{n + 2}^2 \equiv \frac{d r^2}{Z} 
+ r^2 d \Omega_{n + 1}^2 \; .
\end{equation}
Equations $\hat{{\cal R}}_{i j} = \hat{{\cal R}}_{r r} = 0$
imply that the constant parameters $(\hat{a}_i, \hat{c})$ satisfy
the constraints
\begin{equation}\label{achat}
\sum_{i = 0}^{p + 1} \hat{a}_i + n \hat{c} = \frac{1}{2} 
\; , \; \; \; 
\sum_{i = 0}^{p + 1} \hat{a}_i^2 + n \hat{c}^2 
- \hat{c} = \frac{1}{4} \; .
\end{equation}
Throughout in the following we assume that the parameters
$(\hat{a}_i, \hat{c})$ satisfy the constraints given in
(\ref{achat}) and are otherwise arbitrary. Note that besides
$r_0$, there are $p + 3$ parameters $(\hat{a}_i, \hat{c})$, $i =
0, 1, \cdots, p + 1$ satisfying two constraints. Hence, there
are $p + 2$ independent parameters which can be taken to be
$r_0$ and, for example, $\hat{a}_i$, $i = 1, 2, \cdots, p + 1$.

\newpage

\vspace{2ex}

\begin{center}
{\large \bf 3. $D$ dimensional solutions}
\end{center}

\vspace{2ex}

Upon dimensional reduction along $z \equiv X^{p + 1}$ direction,
the solution (\ref{d+1}) gives the $D$ dimensional 
uncharged solution: 
\begin{equation}
d s^2 = - Z^{2 a_0} d t^2 + \sum_{i = 1}^p 
Z^{2 a_i} d X^{i 2} + Z^{2 c} d s_{n + 2}^2 \; , \; \; \; 
e^\phi = Z^q  \; , \; \; \; A_\mu = 0  \label{sch}
\end{equation}
where we assume, with no loss of generality, that $\phi \to 0$
as $r \to \infty$. The parameters $(a_i, c, q)$, $i = 0, 1,
\cdots, p$ are given by
\[
a_i = \hat{a}_i + \frac{\hat{a}_{p + 1}}{D - 2} \; , \; \; \; 
c = \hat{c} + \frac{\hat{a}_{p + 1}}{D - 2} \; , \; \; \; 
q = \frac{\hat{a}_{p + 1}}{b} \; ; \; \; \; 
b = \sqrt{\frac{D - 2}{2 (D - 1)}} 
\]
and, as follows from equations (\ref{achat}), satisfy the
constraints
\begin{equation}\label{acq}
\sum_{i = 0}^p a_i + n c = \frac{1}{2} \; , \; \; \; 
\sum_{i = 0}^p a_i^2 + n c^2 - c + \frac{q^2}{2} 
= \frac{1}{4} \; .
\end{equation}
Throughout in the following we assume that the parameters 
$(a_i, c, q)$ satisfy the constraints given in (\ref{acq}) and
are otherwise arbitrary.

\noindent
Let us note some properties of the solution in (\ref{sch}). 

\noindent {\bf (1)} 
Let $(a_0, c) = (\frac{1}{2}, 0)$. Then necessarily $a_i = q =
0$ for $i \ne 0$. The corresponding solution describes uncharged
$p-$branes in $D$ dimensions with $p$ dimensional compact space;
or, upon a further $p$ dimensional compactification, a $D - p$
dimensional Schwarzschild black hole.

\noindent {\bf (2)} 
Besides $r_0$, there are $p + 3$ parameters $(a_i, c, q)$, $i =
0, 1, \cdots, p$ satisfying two constraints. Hence there are $p
+ 2$ independent parameters. They can be taken to be $r_0$ and,
for example, $(a_i, q)$, $i = 1, 2, \cdots, p$. For $p = 0$, we
get the two parameter solution obtained in \cite{bk}.

\noindent 
{\bf (3)} If $q \ne 0$ then the $D$ dimensional curvature
invariants, for example the Ricci scalar ${\cal R}$, diverge at
$r_0$. Hence, there is a curvature singularity at $r_0$. In $D +
1$ dimensions, however, the Ricci tensor $\hat{R}_{M N}$, and
thus the Ricci Scalar $\hat{R}$ also, vanishes identically
because of the equations of motion. Hence, other curvature
invariants need to be studied to determine the singularities of
the $D + 1$ dimensional solutions. It is important to study the
curvature invariants and the singularities of all the solutions
presented here, but such a study is beyond the scope of the
present paper and will not be pursued here. See \cite{bmo, uses,
glc, bk}, for a discussion of certain issues related to
singularities.

The $D$ dimensional charged solution can be obtained by lifting
the uncharged one to $D + 1$ dimensions, boosting along $z$
direction, and reducing back to $D$ dimensions. Consider a
time-independent $D$ dimensional solution where $A_\mu = 0$,
$e^\phi = e^{\phi_0}$, and the line element is given by
\[
d s^2 = g_{0 0} d t^2 + d s_{D - 1}^2
\]
with $g_{0 \mu} = 0$ for $\mu \ne 0$. This level of
generalisation will suffice for our purposes here. The $D + 1$
dimensional line element is then given by
\[
d \hat{S}^2 = e^{- \frac{2 b \phi_0}{D - 2}} d s^2 
+ e^{2 b \phi_0} d z^2 + \cdots \; 
\equiv \hat{g}_{0 0} d t ^2 
+ \hat{g}_{z z} d z^2 + \cdots 
\]
where $b = \sqrt{\frac{D - 2}{2 (D - 1)}}$, see equation
(\ref{dhattod}). 
Under a boost along $z$ direction
\[
t \to {\cal C} \; t + {\cal S} \; z \; , \; \; \; 
z \to {\cal S} \; t + {\cal C} \; z 
\]
where ${\cal C} = Cosh \; \Theta$, ${\cal S} = Sinh \; \Theta$
with $\Theta$ a boost parameter. The boosted line element 
becomes 
\[
d \hat{S}^2 = H^{- 1} \; \hat{g}_{0 0} d t ^2 
+ H \; \hat{g}_{z z} (d z + A_0 d t)^2 + \cdots 
\]
where $A_0 = \frac{{\cal C} {\cal S} (1 - F)}{H}$ and we have
defined
\begin{equation}\label{hfgen}
H = {\cal C}^2 - F {\cal S}^2 \; \; , \; \; \; 
F = - \hat{g}_{0 0} (\hat{g}_{z z})^{- 1} \; . 
\end{equation} 
Note that $F= - g_{0 0} \; e^{- \lambda \phi}$ where $\lambda =
\frac{2 (D - 1) b}{D - 2}$. Using equation (\ref{dhattod}), the
boosted line element can be reduced to $D$ dimensions. In the
resulting $D$ dimensional charged solution, the non zero
component of the gauge field $A_\mu$ can be written as
\begin{equation}\label{a0}
A_0 = \frac{{\cal C} {\cal S} (1 - F)}{H} 
\simeq - \frac{{\cal S}}{{\cal C}} \; \frac{F}{H} \; , 
\end{equation}
where the last two expressions are physically equivalent since
they differ only by a constant ($= \frac{{\cal S}}{{\cal
C}}$). The line element $d s$ and $e^\phi$ in the charged
solution are given by
\begin{equation}
d s^2 = H^A g_{0 0} d t^2 + H^B d s_{D - 1}^2 
\; , \; \; \; e^\phi = H^C e^{\phi_0} 
\end{equation} 
where $(A, B, C) = \left(- \frac{D - 3}{D - 2}, \frac{1}{D - 2},
\frac{\lambda}{2} \right)$.

Henceforth, unless mentioned otherwise, we assume that
\begin{equation}\label{hf}
H = {\cal C}^2 - F {\cal S}^2 
\; \; , \; \; \; F = Z^{2 k} 
\end{equation}
which is the case for our ansatz (\ref{ansatz}). Then equation
(\ref{a0}) implies that the gauge field charge $Q \propto k \;
{\cal C} {\cal S} \; r_0^n$. Also, in various solutions we
present below, the non zero component of the respective higher
form gauge fields are given by the expression on the right hand
side of the equation (\ref{a0}), and hence the corresponding
charges are also $\propto k \; {\cal C} {\cal S} \; r_0^n$ but
with differing expressions for the parameter $k$, see below.
Therefore, in all the solutions presented here, we will write
the expressions for $d s$, $e^\phi$, and $k$ only.

Thus, the $D$ dimensional charged solution obtained by applying
the boost described above to the solution (\ref{sch}) is given
by
\begin{eqnarray} 
d s^2 & = & - H^A Z^{2 a_0} d t^2 + H^B \left( \sum_{i = 1}^p
Z^{2 a_i} d X^{i 2} + Z^{2 c} d s_{n + 2}^2 \right) \nonumber \\
e^\phi & = & H^C Z^q \; , \; \; \; 
k = a_0 - \frac{\lambda q}{2} \label{d0gen}
\end{eqnarray} 
where $(A, B, C)$ and $\lambda$ are as given above. Note that
in string theory $D = 10$. Then $n = 7 - p$, $b = \frac{2}{3}$,
$\lambda = \frac{3}{2}$, $\; (A, B, C) = (- \frac{7}{8},
\frac{1}{8}, \frac{3}{4})$, and the above solution
becomes
\begin{eqnarray}
d s^2 & = & - H^{- \frac{7}{8}} Z^{2 a_0} d t^2 
+ H^{\frac{1}{8}} \left( \sum_{i = 1}^p Z^{2 a_i} 
d X^{i 2} + Z^{2 c} d s_{9 - p}^2 \right) \nonumber \\
e^\phi & = & H^{\frac{3}{4}}  Z^q \; , \; \; \; 
k = a_0 - \frac{3 q}{4} \; \; . \label{d0}
\end{eqnarray} 
Let us now note some properties of these solutions. 

\noindent {\bf (1)} 
There is now an extra parameter, namely the $(D + 1)$
dimensional boost parameter $\Theta$, which generates the $D$
dimensional charge $Q$. The uncharged solution (\ref{sch}) is
obtained when $\Theta = 0$.

\noindent {\bf (2)} 
Let $(a_0, c) = (\frac{1}{2}, 0)$. Then necessarily $a_i = q =
0$ for $i \ne 0$. Also, $k = \frac{1}{2}$, $F = Z$, and $H =
\left( 1 + \frac{r_0^n {\cal S}^2} {r^n} \right)$. The solution
(\ref{d0gen}) then describes charged $0-$brane smeared in $p$
compact directions. In string theory, $D = 10$ and the solution
(\ref{d0}) describes Dirichlet $0-$brane ($D0-$brane) smeared in
$p$ compact directions.

\noindent {\bf (3)} 
In the extremal limit given by 
\begin{equation}\label{ext}
r_0 \to 0 \; , \; \; \; 
\Theta \to \infty \; , \; \; \; 
r_0^n \; {\cal S}^2 \to finite \; , 
\end{equation}
$Z \to 1$ and hence the values of the parameters $(a_i, c, q)$
are irrelevent, $H \to \left( 1 + \frac{2 k r_0^n {\cal S}^2}
{r^n} \right)$, and the charge $Q \propto k \; {\cal C} {\cal S}
\; r_0^n$ remains finite. The solution (\ref{d0gen}) then
describes extremally charged $0-$brane smeared in $p$ compact
directions. In string theory, $D = 10$ and the solution
(\ref{d0}) then describes extremal $D0-$brane smeared in $p$
compact directions.

\vspace{2ex}

\begin{center}
{\large \bf 4. A few brane solutions of string/M theories} 
\end{center}

\vspace{2ex}

Starting from the $D0-$brane solutions \footnote{ Strictly
speaking, equation (\ref{d0}) is not a $D0-$brane solution if
$(a_0, c) \ne (\frac{1}{2}, 0)$. Nevertheless, even then we will
continue to use such phrases to refer to the solutions.} given
in (\ref{d0}), one may obtain other brane solutions of string/M
theories by repeated use of $S$ and $T$ dualities and boosts in
$11^{th}$ dimension \cite{rules, boost}. We present a few
examples below.

The boost in $11^{th}$ dimension is as given before but now with
$D = 10$. The transformation rules for S and T duality are given
in \cite{bho}. For the cases of interest here, they are simple
enough and we will now explain these transformations briefly as
applied to $d s$ and $e^\phi$.

Under S duality, $(g_{\mu \nu}, \phi) \to (g_{\mu \nu}, -
\phi)$. The T duality rules are given in \cite{bho} in string
frame where the string metric $G_{\mu \nu} = e^{\frac{\phi}{2}}
\; g_{\mu \nu}$. One may convert the Einstein frame solutions to
the string frame, apply T duality rules, and convert back to the
Einstein frame. The result of a T duality, along {\em e.g.}
$X^1$ direction, applied to the $D0-$brane solution (\ref{d0})
is that the original $1-$form gauge field $A_0$ now becomes a
$2-$form gauge field $A_{0 1}$ and
\begin{eqnarray} 
d s^2 & = & H^{- \frac{6}{8}} \left(- Z^{2 \bar{a}_0} d t^2 
+ Z^{2 \bar{a}_1} d X^{1 2} \right) + H^{\frac{2}{8}} 
\left( \sum_{i = 2}^p Z^{2 \bar{a}_i} d X^{i 2} 
+ Z^{2 \bar{c}} d s_{9 - p}^2 \right) \nonumber \\ 
e^\phi& = & H^{\frac{2}{4}} Z^{\bar{q}} \; , \; \; \; 
k = a_0 - \frac{3 q}{4} = \bar{a}_0 + \bar{a}_1 
- \frac{2 \bar{q}}{4}  \label{d1} 
\end{eqnarray}  
where $(\bar{a}_i, \bar{c}, \bar{q}; \bar{a}_1) = (a_i, c, q;
a_1) \; + \; \left(\frac{1}{4}, \frac{1}{4}, - 1; - \frac{7}{4}
\right) \chi$ with $\chi = a_1 + \frac{q}{4}$ and $i \ne 1$.

Note that after a T duality, the parameters $(a_i, c, q)$ have
changed to $(\bar{a}_i, \bar{c}, \bar{q})$, $i = 0, 1, \cdots,
p$. Also, as can be verified easily, if $(a_i, c, q)$ satisfy
the constraints (\ref{acq}) then so do $(\bar{a}_i, \bar{c},
\bar{q})$. However, $k$ has not changed. But when expressed in
terms of $(\bar{a}_i, \bar{c}, \bar{q})$, it has a different
form.

This will be the case in general in the following. However,
instead of ornamenting the labels $(a_i, c, q)$ endlessly after
each $S$ or $T$ duality, we will simply use the same labels
$(a_i, c, q)$ always. Consequently, of course, the expression
for $k$ will be different. Similarly, for $M$ theory, we will
always use the labels $(\hat{a}_i, \hat{c})$. Note that $(a_i,
c, q)$ will always satisfy the constraints given in (\ref{acq}),
and $(\hat{a}_i, \hat{c})$ those given in (\ref{achat}), but are
otherwise arbitrary.

Consider now T duality along $X^i$ directions, $i = 1, 2,
\cdots, p_1 \le p$ in the $D0-$brane solution given in
(\ref{d0}). The calculation is straightforward and the result is
that the original 1-form gauge field $A_0$ now becomes 
a ($p_1 + 1$)-form gauge field $A_{0 1 \cdots p_1}$ and
\begin{eqnarray}
d s^2 & =  & H^A \left(- Z^{2 a_0} d t^2 
+ \sum_{i = 1}^{p_1} Z^{2 a_i} d X^{i 2} \right) 
+ H^B \left( \sum_{p_1 + 1}^p Z^{2 a_i} d X^{i 2} 
+ Z^{2 c} d s_{9 - p}^2 \right)\nonumber \\
e^\phi & = & H^C Z^q \; , \; \; \; 
k = \sum_{i = 0}^{p_1} a_i 
- \frac{(3 - p_1) q}{4} \label{dp1} 
\end{eqnarray} 
where $(A, B, C) = (- \frac{7 - p_1}{8}, \frac{p_1 + 1}{8},
\frac{3 - p_1}{4})$. Equation (\ref{dp1}) describes
$Dp_1-$branes smeared in the remaining $(p - p_1)$ compact
directions. We will now write down explicitly a few brane
solutions of string/M theories.

\noindent
$Dp-$branes are obtained by setting $p_1 = p$ in equation
(\ref{dp1}):
\begin{eqnarray}
d s_{Dp}^2 & =  & H^{\frac{p - 7}{8}} \left(- Z^{2 a_0} 
d t^2 + \sum_{i = 1}^p Z^{2 a_i} d X^{i 2} \right) 
+ H^{\frac{p + 1}{8}} Z^{2 c} d s_{9 - p}^2  \nonumber \\ 
e^\phi & = & H^{\frac{3 - p}{4}} Z^q \; , \; \; \; 
k = \sum_{i = 0}^p a_i - \frac{(3 - p) q}{4} \; . \label{dp} 
\end{eqnarray} 

\noindent
Fundamental (F) strings are obtained by S dualising $D1-$branes: 
\begin{eqnarray}
d s_F^2 & =  & H^{- \frac{3}{4}} \left(- Z^{2 a_0} d t^2 
+ Z^{2 a_1} d X^{1 2} \right) 
+ H^{\frac{1}{4}} Z^{2 c} d s_8^2  \nonumber \\
e^\phi & = & H^{- \frac{1}{2}} Z^q \; , \; \; \; 
k = \sum_{i = 0}^1 a_i + \frac{q}{2} \; . \label{f} 
\end{eqnarray} 

\noindent
NS sector $5-$branes are obtained by S dualising $D5-$branes:
\begin{eqnarray}
d s_5^2 & =  & H^{- \frac{1}{4}} \left(- Z^{2 a_0} d t^2 
+ \sum_{i = 1}^5 Z^{2 a_i} d X^{i 2} \right) 
+ H^{\frac{3}{4}} Z^{2 c} d s_4^2 \nonumber \\
e^\phi & = & H^{\frac{1}{2}} Z^q \; , \; \; \; 
k = \sum_{i = 0}^5 a_i - \frac{q}{2} \; . \label{ns5} 
\end{eqnarray} 

\noindent
$M2-$branes are obtained by lifting F strings to 11 dimensions:
\begin{equation}
d s_2^2 = H^{- \frac{2}{3}} \left(
- Z^{2 \hat{a}_0} d t^2 + \sum_{i = 1}^2 Z^{2 \hat{a}_i} 
d X^{i 2} \right) + H^{\frac{1}{3}} Z^{2 \hat{c}} d s_8^2 
\; , \; \; \; 
k = \sum_{i = 0}^2 \hat{a}_i \; . \label{m2} 
\end{equation} 

\noindent
$M5-$branes are obtained by lifting $D4-$branes to 11
dimensions:
\begin{equation}
d s_5^2 = H^{- \frac{1}{3}} \left(
- Z^{2 \hat{a}_0} d t^2 + \sum_{i = 1}^5 Z^{2 \hat{a}_i} 
d X^{i 2} \right) + H^{\frac{2}{3}} Z^{2 \hat{c}} d s_5^2 
\; , \; \; \; 
k = \sum_{i = 0}^5 \hat{a}_i \; . \label{m5} 
\end{equation} 

Intersecting brane solutions can also be obtained by further
boosts. As an example, we now obtain intersecting $D1-D5$ brane
solution. Start with the $D4-$brane solution smeared in a
compact $X^5$ direction, {\em i.e.} the solution (\ref{dp1})
with $p_1 = 4$ and $p = 5$, lift it to 11 dimension, boost along
the $11^{th}$ direction which will generate a second charge, and
reduce back to 10 dimensions. This will now give the $D0-D4$
branes smeared in the $X^5$ direction:
\begin{eqnarray} 
d s^2 & = & - H^{- \frac{3}{8}} h^{- \frac{7}{8}} 
Z^{2 a_0} d t^2 + H^{- \frac{3}{8}} h^{\frac{1}{8}} 
\sum_{i = 1}^4 Z^{2 a_i} d X^{i 2} \nonumber \\ 
& + & H^{\frac{5}{8}} h^{\frac{1}{8}} \left( Z^{2 a_5} 
d X^{5 2} + Z^{2 c} d s_4^2 \right) \nonumber \\ 
e^\phi & = & H^{- \frac{1}{4}} h^{\frac{3}{4}} Z^q 
\; , \; \; \; k = \sum_{i = 0}^4 a_i + \frac{q}{4} 
\nonumber \\ 
h & = & {\cal C}_1^2 - f {\cal S}_1^2 \; \; , \; \; \; 
f = Z^{2 l} \; \; , \; \; \; 
l = a_0 - \frac{3 q}{4} \label{d0d4} 
\end{eqnarray} 
where ${\cal C}_1 = Cosh \; \Theta_1$, ${\cal S}_1 = Sinh \;
\Theta_1$ with $\Theta_1$ a boost parameter that generates the
$D0-$brane charge. The $D0-$brane gauge field is given by a
similar expression as in (\ref{a0}) and (\ref{hf}), but with
$(H, F, k)$ replaced by $(h, f, l)$.

\noindent
$D1-D5$ intersecting branes are obtained by T dualising the
$D0-D4$ branes above along the $X^5$ direction:
\begin{eqnarray}
d s^2 & =  & H^{- \frac{2}{8}} h^{- \frac{6}{8}} 
\left( - Z^{2 a_0} d t^2 + Z^{2 a_5} d X^{5 2} \right) 
\nonumber \\
& + & H^{- \frac{2}{8}} h^{\frac{2}{8}} 
\sum_{i = 1}^4 Z^{2 a_i} d X^{i 2} + H^{\frac{6}{8}} 
h^{\frac{2}{8}} Z^{2 c} d s_4^2  \nonumber \\
e^\phi & = & H^{- \frac{1}{2}} h^{\frac{1}{2}} Z^q \; , \; \; \; 
k = \sum_{i = 0}^5 a_i + \frac{q}{2} 
\; \; , \; \; \; l = a_0 + a_5 - \frac{q}{2} \label{d1d5} 
\end{eqnarray} 
where $h$ and $f$ are given in (\ref{d0d4}) but now with $l$
given as above. 

\noindent
Let us now note some properties of these solutions. 

\noindent {\bf (1)} 
The parameters in all the above solutions are $r_0$, the boost
parameters, and $(a_i, c, q)$ or $(\hat{a}_i, \hat{c})$, $i = 0,
1, \cdots$ satisfying two constraints.

\noindent {\bf (2)} 
The extremal limit is given as in (\ref{ext}) for each boost
parameter $\Theta$. 

\noindent {\bf (3)} 
One can further boost the $D1-D5$ solution in (\ref{d1d5}) along
the common isometric direction $X^5$. This will give a three
charged system which can also be converted to three intersecting
brane systems by a chain of S, T, U dualities. The corresponding
solutions for this and other intersecting brane configurations
in string/M theories can all be obtained striaghtforwardly by
suitable combinations of lifting to $11$ dimensions, boosts, S,
and T dualities.

\noindent 
{\bf (4)} The intersecting brane solutions for general values of
$r_0$, the boost parameters, and $(a_i, c, q)$ or $(\hat{a}_i,
\hat{c})$ can be obtained from the corresponding extremal ones
by simple rules, similar to those given in \cite{rules} for $a_i
= c = q = 0$, $i = 1, 2, \cdots$. These rules are applicable
here also with minor differences: In the present case, the non
compact transverse space has a line element $d s_{n + 2}$ given
in (\ref{dsn+2}). One inserts $(Z^{2 a_i}, Z^{2 c}, Z^q)$
factors in the appropriate places. The parameters $(a_i, c, q)$,
or $(\hat{a}_i, \hat{c})$, satisfy the contsraints given in
(\ref{acq}), or (\ref{achat}), but are otherwise arbitrary. The
functions $H$ and $F$ associated with each brane are of the form
(\ref{hf}), with the exponent $k$ given by
\[
k = \sum_{i \in brane} a_i - \frac{\lambda_p q}{2}
\]
where $i \in brane$ means that $i$ in the summation runs over
the world volume indices of the corresponding brane, and
$\lambda_p = 0, - 1, + 1, \frac{3 - p}{2}$ for M theory branes,
F strings, NS $5-$branes, and $Dp-$branes respectively. If there
is a common isometric direction along which branes intersect
then a further boost can be added in that direction, see {\bf
(3)} above. The corresponding $H$ and $F$ can be obtained using
a formula similar to (\ref{hfgen}). See \cite{rules} for more
details.

\vspace{2ex}

\begin{center}
{\large \bf 5. Brane solutions and a duality property}

\end{center}

\vspace{2ex}

Let the spacetime be D dimensional with $p$ compact directions.
Consider the equations of motion that follow from the action
given in (\ref{sp}) where $\lambda_{p_1}$ is an arbitrary
constant which may depend on $p_1$. Their solutions that
describe electric type $p_1-$branes smeared over $(p - p_1)$
compact directions are given by
\begin{eqnarray}
d s^2 & = & H^A \left(- Z^{2 a_0} d t^2 
+ \sum_{i = 1}^{p_1} Z^{2 a_i} d X^{i 2} \right) \nonumber \\
& + & H^B \left( \sum_{i = p_1 + 1}^p Z^{2 a_i} d X^{i 2} 
+ Z^{2 c} d s_{n + 2}^2 \right)\nonumber \\
e^\phi & = & H^C Z^q \; , \; \; \; 
k = \sum_{i = 0}^{p_1} a_i 
- \frac{\lambda_{p_1} q}{2} \nonumber \\ 
A_{0 1 \cdots p_1} & = & \frac{{\cal C} {\cal S} (1 - F)}{H} 
\simeq - \frac{{\cal S}}{{\cal C}} \; \frac{F}{H} 
\label{gen} 
\end{eqnarray} 
where $(A, B, C) = \left( - \; \frac{2 (D - 3 - p_1)}{\Delta} ,
\; \frac{2 (p_1 + 1)} {\Delta} , \; \frac{(D - 2) \lambda_{p_1}}
{\Delta} \right)$, other symbols are all as defined before, and
\begin{equation}\label{delta}
\Delta = (p_1 + 1) (D - 3 - p_1) 
+ \frac{(D - 2) \lambda^2_{p_1}}{2} \; .
\end{equation}
We obtained the solutions (\ref{gen}) by applying the rules
given in {\bf (4)}, below equation (\ref{d1d5}), to the extremal
solutions in \cite{solns, glc} and then verified that the
equations of motion are satisfied. Instead of presenting these
details of verification, we will now discuss some properties of
the solutions (\ref{gen}) and then relate them to the
corresponding solutions in \cite{mo} which were obtained by
solving the equations of motion directly.

The ADM mass $M$ for the solutions, such as those in
(\ref{gen}), is defined by \cite{solns,bmo}
\begin{equation}\label{m}
\lim_{R \to \infty}  \; g_{0 0} = - 1 
+ \frac{2 \kappa^2 \; M}{(n + 1) \omega_{n + 1} V_p R^n}
\end{equation}
where $V_p$ is the volume of the $p$ dimensional compact space,
$\omega_{n + 1}$ is the area of the $(n + 1)$ dimensional unit
sphere, $\kappa^2$ is given in (\ref{sp}), and $R$ is the
isotropic coordinate defined by
\begin{equation}\label{Rdefn}
d s_{n + 2}^2 = \frac{d r^2}{Z} + r^2 d \Omega_{n + 1}^2 
= {\cal Z} \; \left( d R^2 + R^2 d \Omega_{n + 1}^2 \right) \; .
\end{equation}
Equation (\ref{Rdefn}) implies that ${\cal Z} = \frac{r^2}{R^2}$
and $\frac{d R}{R} = \frac{d r}{r \sqrt{Z}}$, from which the
functions $R(r)$ and $r(R)$ can be easily obtained upto a
constant factor. Fixing this factor by requiring $R \to r$ as 
$r \to \infty$, we get
\begin{equation}\label{iso}
2 R^n = r^n - \frac{r_0^n}{2} + r^n \; 
\sqrt{1 - \frac{r_0^n}{r^n}} \; , \; \; \; 
r^n = R^n \; (1 + Y)^2 
\end{equation} 
where we have defined $Y = \frac{R_0^n}{R^n}$ and $R_0 =
R(r_0)$; hence $R_0^n = \frac{r_0^n}{4}$. Also, $Z = G^2$ where
\begin{equation}\label{G}
G = \frac{1 - Y}{1 + Y} \; . 
\end{equation}

Substituting these expressions into the solutions given in
(\ref{gen}) and using the expressions for $H$ and $F$ given in
(\ref{hf}), it follows that the ADM mass $M$ defined in
(\ref{m}) is given by
\begin{equation}\label{mgen}
M = {\cal N} \left( a_0 - k A {\cal S}^2 \right) r_0^n 
\end{equation}
where ${\cal N} = \frac{(n + 1) \omega_{n + 1} V_p}{\kappa^2}$.
Physically, the mass $M$ must be positive which will impose a
mild inequality on the parameters of the solution.

The brane charge $Q$ of the solutions (\ref{gen}) can also be
defined similarly using the asymptotic behaviour of the
corresponding gauge potential. It is clear that $Q \propto k \;
{\cal C} {\cal S} \; r_0^n$. We define the proportionality
constant here so that we get $\vert Q \vert = M$ in the extremal
limit (\ref{ext}). Hence,
\begin{equation}\label{qgen}
Q = \pm \; {\cal N} A \; k {\cal C} {\cal S} \; r_0^n \; .
\end{equation}

We will now discuss a duality property of the solutions
(\ref{gen}). First, note that if $D = 10$ and $\lambda_p =
\frac{3 - p}{2}$ then $\Delta = 16$ and equations (\ref{gen})
reduce to the $Dp_1-$brane solutions (\ref{dp1}) from which
other Dirichlet branes can be obtained by T dualities. It turns
out that the solutions (\ref{gen}) also have a similar duality
property which is valid for any value of $D$ if $\lambda_p$ is a
specific function of $p$ to be obtained below. We now describe
this duality as applied to the solution (\ref{gen}).

Consider a ``$w-$frame'' where the metric $G_{\mu \nu}$ is
given by
\begin{equation}\label{wframe}
G_{\mu \nu} = e^{w \phi} \; g_{\mu \nu}  
\end{equation}
with $w$ a constant. Now, in the context of the $p_1-$brane
solutions (\ref{gen}), consider the duality along an isometric
direction $X^j$, $j \in (1, 2, \cdots,p)$. We denote this
duality by $T_j$ and define it in the $w-$frame by the following
transformations:
\begin{equation}\label{uv}
e^\phi \to e^{\bar{\phi}} =  G_{j j}^{- u} 
\; e^{v \; \phi}  \; , \; \; \;  
G_{j j} \to \bar{G}_{j j} = G_{j j}^{- U} 
\; e^{V \phi} 
\end{equation}
where $(u, v, U, V)$ are constants; the $(p_1 + 1)$ form gauge
field becomes a $p_1$ or a $(p_1 + 2)$ form gauge field
according to whether $X^j$ is parellel or transverse to the
$p_1-$brane worldvolume respectively; and other fields remain
unchanged in the $w-$frame. 

Let $T_j$ and $T_{j'}$, $j \ne j'$, be the dualities (\ref{uv})
along two isometric directions $X^j$ and $X^{j'}$ respectively.
We now impose the following natural requirements on the duality
(\ref{uv}), in close analogy with T duality of the string
theory. 

\vspace{2ex}

\noindent
{\bf (I)} $\; $ $T_j^2 = I$, the identity. Then $U = v$
and $u V = v^2 - 1$. 

\vspace{2ex}

\noindent
{\bf (II)} $\; $ $T_j T_{j'} = T_{j'} T_j$. Then $v = 1$ and
$u V = 0$. 

\vspace{2ex}

\noindent
{\bf (III)} $\;$ Under (\ref{uv}), the $p_1-$brane solutions
(\ref{gen}) transform to a $(p_1 - 1)$ or a $(p_1 + 1)$ brane
solutions according to whether $X^j$ is parellel or transverse
to the $p_1-$brane worldvolume respectively. After the duality,
$X^j$ becomes transverse or parallel to the $(p_1 -1)$ or the
$(p_1 + 1)$ brane respectively. 

\vspace{2ex}

Let the parameters $(a_i, c, q)$, $(A, B, C)$, and $A_j$ defined
by $g_{j j} \equiv H^{A_j} Z^{2 a_j}$ in the solution
(\ref{gen}) transform under the duality (\ref{uv}) to
$(\bar{a}_i, \bar{c}, \bar{q})$, $(\bar{A}, \bar{B}, \bar{C})$,
and $\bar{A_j}$ respectively. \footnote{ $(\bar{a}_i, \bar{c},
\bar{q})$ and $(\bar{A}, \bar{B}, \bar{C}; \bar{A_j})$ can be
obtained by converting the solutions (\ref{gen}) to the
$w-$frame, applying duality (\ref{uv}), and converting back to
the Einstein frame. The resulting expressions, valid for any
value of $(u, v, U, V)$, are unwieldy and will not be
presented.} The meaning of {\bf (III)} and its consequences,
obtained after some algebra, can now be stated as follows.

\noindent
{\bf (a)} The parameters $(\bar{a}_i, \bar{c}, \bar{q})$ must
satisfy the constraints given in (\ref{acq}). This implies, upon
using $U = v = 1$ and $u V = 0$ and after some algebra, that
\begin{equation}\label{v}
(u, v, U, V) = (w, 1, 1, 0) \; , \; \; \; 
w^2 = \frac{2}{D - 2} \; .
\end{equation}
The $T_j$ duality transformation (\ref{uv}) is now similar to
that of T duality of the string theory:
\begin{equation}\label{v=1}
e^{\bar{\phi}} =  G_{j j}^{- w} \; e^{\phi}  \; , \; \; \;  
G_{j j} \to \bar{G}_{j j} = G_{j j}^{- 1} \; , 
\end{equation}
under which the transformed parameters $(\bar{a}_i, \bar{c},
\bar{q}; \bar{a}_j)$, $i \ne j$ and $(\bar{A}, \bar{B}, \bar{C};
\bar{A_j})$ are given by
\begin{eqnarray}
(\bar{a}_i, \bar{c}, \bar{q}; \bar{a}_j) & = & (a_i, c, q; a_j)
\; \; \; + (w^2, w^2, - 2 w; w^2 - 2) \; \chi \nonumber \\
(\bar{A}, \bar{B}, \bar{C}; \bar{A_j}) & = & (A, B, C; A_j) 
+ (w^2, w^2, - w \; ; w^2 - 2) \; \psi \label{aA}
\end{eqnarray} 
where $\chi = a _j + \frac{w q}{2}$ and $\psi = A_j + w C$.

\noindent
{\bf (b)} If $X^j$ is parallel or transverse to the $p_1-$brane
worldvolume then one must have 
\begin{eqnarray}
& & A_j = A \; , \; \; \; 
\bar{A_j} = \bar{B} \; \; , \; \; \; (\bar{A}, \bar{B}, 
\bar{C}) = (A, B, C) \vert_{p_1 - 1} \nonumber \\
& or  \; \; \; & A_j = B \; , \; \; \; 
\bar{A_j} = \bar{A} \; \; , \; \; \; (\bar{A}, 
\bar{B}, \bar{C}) = (A, B, C) \vert_{p_1 + 1} \label{aj}
\end{eqnarray}
respectively where $(\bar{A}, \bar{B}, \bar{C}; \bar{A_j})$ are
given in (\ref{aA}) and $(A, B, C) \vert_{p_1 \pm 1}$ means that
$(A, B, C)$ are as given below equation (\ref{gen}) but with
$p_1$ replaced by $p_1 \pm 1$.

Equations $\bar{A_j} = \bar{B}$ or $= \bar{A}$ imply that $A + B
+ 2 w C = 0$ which in turn implies that $\lambda_{p_1}$ must be
a specific function of $p_1$: $\lambda_{p_1} = \frac{(D - 4 - 2
p_1) w}{2}$. Using the expressions for $(A, B, C)$, $\lambda_p$,
and $w^2$, we get $A + w C = - (B + w C)= - \frac{D -
2}{\Delta}$. Using the transformations in (\ref{aA}), it is now
straightforward to show that the remaining equations in
(\ref{aj}) are satisfied. Note also that, with $\lambda_p$ given
as above, we now have $\Delta = \frac{(D - 2)^2}{4}$ as follows
from equation (\ref{delta}).

\noindent
{\bf (c)} The expression for $k$ in (\ref{gen}) does not
transform. But, it must have the correct form when written in
terms of $(\bar{a}_i, \bar{c}, \bar{q})$ given in (\ref{aA}).
That is, if we set, with no loss of generality, $j = p_1$ or $=
p_1 + 1$ when $X^j$ is parallel or transverse to the brane
respectively then the expression $k = \sum_{i = 0}^{p_1} a_i -
\frac{\lambda_{p_1} q}{2}$ must also be expressible respectively
as
\begin{equation}\label{kbar}
k = \sum_{i = 0}^{p_1 - 1} \bar{a}_i 
- \frac{\lambda_{p_1 - 1} \bar{q}}{2}  
\; \; \; \; \; \; or \; \; \; \; \; \; 
=  \sum_{i = 0}^{p_1 + 1} \bar{a}_i 
- \frac{\lambda_{p_1 + 1} \bar{q}}{2} \; \; . 
\end{equation}

Using the transformations in (\ref{aA}), the specific function
for $\lambda_p$ obtained above, and $w^2 = \frac{2}{D - 2}$, it
is straightforward to show that equations (\ref{kbar}) are
satisfied. From now on in the following we assume that
\begin{equation}\label{w}
\lambda_{p_1} = \frac{D - 4 - 2 p_1}{2} \; w 
\; \; , \; \; \; w = \sqrt{\frac{2}{D - 2}}
\end{equation}
where we have chosen a positive sign for $w$ with no loss of
generality. 

We now make a few remarks about the duality described above.

\noindent
{\bf (1)} The $D$ dimensional action (\ref{sp}) written in the
$w-$frame (\ref{wframe}) becomes
\begin{equation}\label{spwframe}
S_{p_1} = \frac{1}{2 \kappa^2} \int d^D X \sqrt{- G} 
\left\{ e^{- \frac{\phi}{w}} \left(
{\cal R}_G + \frac{1}{w^2} (\partial \phi)^2 \right)
- \frac{ F_{p_1 + 2}^2}{2 (p_1 + 2) !}  \right \} \; . 
\end{equation}
Note that the function $\lambda_{p_1}$ is such that the
$p_1-$brane field strength $F_{p_1 + 2}$ does not couple to
$\phi$ in the $w-$frame.

\noindent
{\bf (2)} The duality (\ref{v=1}) is analogous to T duality of
string theory but is valid for any value of $D$. For $D = 10$,
and for $D < 10$ also, this is same as T duality of the string
theory and all the expressions above reduce to the well known
ones. For $D > 10$, this duality can be used as a solution
generating technique but otherwise its significance, if any, is
not clear.

\noindent
{\bf (3)} The $0-$brane solutions in (\ref{gen}) can be
generated by a $(D + 1)$ dimensional boost iff $\lambda_0$ in
(\ref{w}) equals $\lambda$ in (\ref{s}) which is possible only
for $D = 10$. If $D \ne 10$ then it is not clear how to generate
the $0-$brane solutions without actually solving the relevent
equations of motion.

\noindent
{\bf (4)} Note that the duality symmetry (\ref{v=1}) transforms
$p_1-$branes into $(p_1 \pm 1)-$branes and is not a symmetry of
the equations of motion following from action (\ref{sp}). With
no gauge fields, however, it is indeed a symmetry of the
equations of motion. Just as in T duality of the string theory,
this is a symmetry between long and short distances in the
compact directions since $G_{j j} \to \frac{1}{G_{j j}}$. For
the standard solutions where $(a_0, c) = (\frac{1}{2}, 0)$, and
hence $a_i = q = 0$ for $i \ne 0$, the duality transformations
are trivial. It may be of interest to consider other solutions
presented here and explore the consequences of this duality. See
\cite{cosm} where a similar duality in time dependent solutions
is explored in detail.

\vspace{2ex}

\begin{center}
{\large \bf 6. Conclusion}
\end{center}

\vspace{2ex}

In this paper we have shown that the multiparameter brane
solutions of string/M theories given in the literature can all
be obtained by a suitable combination of boosts in eleven
dimension, S and T dualities. We also described a duality
property of the $D$ dimensional multiparameter solutions
describing branes smeared in compact directions. 

We have not discussed any physical implications of these
solutions since they are beyond the scope of this paper. But
they are likely to be interesting and it is important to study
them. We conclude now by mentioning a few aspects of the present
solutions that may be studied further.

One should understand the brane antibrane interpretations \cite{
bmo, kam, uses, lr} of such solutions in terms of the present
parametrisations and see if any insights can be obtained into
tachyon condenstaion and the related dynamics. Multiparameter
solutions have also been applied to the study of the so called S
branes \cite{sbrane, lr}. It is of interest to study whether the
present parametrisation in terms of boosts, S and T dualities
applies in that context also.

Also, one should understand the singularities of the
multiparameter solutions, their physical relevence and
implications, and their resolution if possible. Perhaps, these
can be studied along the lines of \cite{bmo, uses, glc, bk}. It
may also be of interest to study the multiparameter solutions
using various string/M theory branes as probes and studying the
geodesic motions of such probes. Our preliminary calculations,
and also the results of \cite{bk}, indicate that such a study is
likely to be fruitful.

\vspace{2ex}

\begin{center}
{\large \bf Appendix: Relation to other solutions}
\end{center}

\vspace{2ex}

Recently, multiparameter brane solutions have been obtained by
solving directly the equations of motion at various levels of
generality \cite{zz, bmo, kam, glc, lr, mo}. These multi
parameter solutions, with suitable symmetry such as $SO(p)$ or
$ISO(p, 1)$, are interpreted to represent non BPS brane
antibrane systems \cite{bmo, kam, glc, lr, mo}. The
singularities of such solutions have been discussed in
\cite{bmo, uses, glc}. The most general multiparameter brane
solutions are given by Miao and Ohta (MO) in \cite{mo} from
which all the others in \cite{zz, bmo, kam, glc, lr} can be
obtained.

The (intersecting) brane solutions for $D = 10, 11$, namely for
string/M theories, given by MO in \cite{mo} can all be obtained
by the method presented here by repeated use of boosts, T and S
dualities on the solution (\ref{sch}). Also, the smeared
$p_1-$brane solutions of \cite{mo} for any value of $D$ are
indeed the same as the present general solutions (\ref{gen}),
obtained here by applying the rules given in {\bf (4)} below
equation (\ref{d1d5}) to the extremal solutions in \cite{solns,
glc}. This can be shown as follows.

Consider the MO solutions given in \cite{mo}. The main equations
required are those numbered (6, 26, 27, 29, 33-35, 38, 39,
42-47) in \cite{mo}. Their $\tilde{d} = n$ and their $\frac{1}{x
\mu} \; (c_0, c_\alpha, c_b, c_\phi) = (a_0, a_i, c - \frac{1}{2
n}, q)$ where $x \equiv \sqrt{2 \nu - 1} \ge 0$ and $\mu, \nu$
are constant parameters in \cite{mo}. Then, their constraints
(42) and (29) are the same as those in (\ref{acq}) here. Also,
their $\rho = \frac{x - 1} {x + 1}$. Consider now their $(n +
2)$ dimensional transverse line element given by
\[
d S_{n + 2}^2 \vert_{MO} = \left( \frac{f X^2}{g^2} 
\right)^{\frac{1}{n}} \; \left( \frac{d \tilde{r}^2}{f} 
+ \tilde{r}^2 d \Omega_{n + 1}^2  \right) 
\]
where the functions $g$, $X$, and $f$ can be written, after some
algebra, as $g = \vert \tilde{g} \vert$,
\[
\tilde{g} = \frac{\sqrt{f} - \rho}{1 - \rho \sqrt{f}} 
\; , \; \; \; 
X = \frac{(\sqrt{f} - \rho) \; ({1 - \rho \sqrt{f}})}
{(1 - \rho)^2 \; \sqrt{f}} \; , \; \; \; 
f = 1 - \frac{\mu}{\tilde{r}^n} \; . 
\]
Define the isotropic coordinate $R$ by 
\[
\frac{d \tilde{r}^2}{f} + \tilde{r}^2 d \Omega_{n + 1}^2 = 
{\cal F} \; \left( d R^2 + R^2 d \Omega_{n + 1}^2 \right) \; .
\]
It then follows that $2 R^n = \tilde{r}^n - \frac{\mu}{2} +
\tilde{r}^n \; \sqrt{1 - \frac{\mu}{\tilde{r}^n}} \; , \; $
${\cal F} = \frac{\tilde{r}^2}{R^2} = (1 + Y_1)^{\frac{4}{n}}$,
and $f = \left( \frac{1 - Y_1}{1 + Y_1} \right)^2$ where $Y_1 =
\frac{\mu}{4 R^n}$. Using these expressions, we get
\[
d S_{n + 2}^2 \vert_{MO} = \left( \frac{f X^2}{g^2} \; 
(1 + Y_1)^4 \right)^{\frac{1}{n}} \; \left( 
d R^2 + R^2 d \Omega_{n + 1}^2 \right) 
\]
Now, define $r_0^n = x \mu$, $R_0^n = \frac{r_0^n}{4}$, and $Y =
x Y_1 = \frac{R_0^n}{R^n}$. The expressions for $\tilde{g}$ and
$X \sqrt{f}$ now simplify considerably and, after some algebra,
are given by
\[
\tilde{g} = \frac{1 - Y}{1 + Y} 
\; , \; \; \; X \sqrt{f} 
= \frac{1 - Y^2}{(1 + Y_1)^2} \; . 
\]
Using these expressions, it follows that 
\[
d S_{n + 2}^2 \vert_{MO} = (1 + Y)^{\frac{4}{n}} \;
\left( d R^2 + R^2 d \Omega_{n + 1}^2 \right) \; .
\]
Also, with $r_0^n = x \mu$, one gets $g^2 = \tilde{g}^2 = \left(
\frac{1 - Y}{1 + Y} \right)^2 = 1 - \frac{r_0^n}{r_n} =
Z$. Hence, with no loss of generality, we take $g = G$ given in
(\ref{G}).

Remarkably $Y_1$, and hence $\mu$, has disappeared completely
from $\tilde{g}$ and the line element $d S_{n + 2}^2 \vert_{MO}$
and only $Y$, and hence the combination $x \mu$, remains. This is
now true for all the solutions given in \cite{mo} as can be seen
easily. This shows that MO solutions do not depend on two
independent parameters $\mu$ and $\nu$, but only on one
combination given by $r_0^n = x \mu = \mu \; \sqrt{2 \nu - 1}$.
\footnote{
The relation between the radial coordinate $\tilde{r}$ of MO and
the present one $r$ is given by
\[
2 R^n = \tilde{r}^n - \frac{\mu}{2} + \tilde{r}^n \; 
\sqrt{1 - \frac{\mu}{\tilde{r}^n}}
= r^n - \frac{x \mu}{2} + r^n \; 
\sqrt{1 - \frac{x \mu}{r^n}} 
\]
which, indeed, involves both the parameters $\mu$ and $\nu$. But
this is just a diffeomorphism. Only the combination $x \mu$ that
appear in the solutions is physically relevent.} 

The expressions for the gauge field in \cite{mo} can now be
written in terms of $Z$. Putting together various expressions
and setting their $\beta = {\cal C}^2$ here, it can be seen
straightforwardly that their gauge field matches the present one
given in (\ref{a0}) including the exponent $k$. It can now be
seen by using the above expressions in the MO solutions that the
(intersecting) brane solutions for $D = 10, 11$, namely for
string/M theories, given by MO in \cite{mo} can all be obtained
by the method presented here by repeated use of boosts, S and T
dualities on the solution (\ref{sch}). Similarly, it can also be
seen that the smeared $p_1-$brane solutions of \cite{mo} for any
value of $D$, namely equation (44) with $A = 1$ in \cite{mo},
are indeed the same as the general solutions (\ref{gen}) here.

In \cite{mo} Miao and Ohta also show that the solutions given in
\cite{zz, bmo, kam}, which we refer to as ZZ ones, follow from
the MO solutions; hence they follow from the present solutions
also. Alternately, this can be shown directly by relating
(\ref{gen}), with $p_1 = p$, to the ZZ solutions. The steps are
straightforward but involve a fair amount of tedious algebra. We
now mention relations between a few select quantities in the ZZ
solutions and the present ones.

The ZZ solutions have $SO(p)$ symmetry, hence set $a_1 = a_2 =
\cdots = a_p \equiv a$ here; $ISO(p, 1)$ symmetry requires
further that $a_0 = a$.  Also, their $r$ coordinate is our
isotropic coordinate $R$, their $e^h = G$ here, and their $Cosh
\; (k_{ZZ} h) - c_2 \; Sinh \; (k_{ZZ} h) = G^{- k_{ZZ}} \; H$
where $c_2 = {\cal C}^2 + {\cal S}^2$, $k_{ZZ} = 2 k = 2 a_0 + 2
p a - \lambda_p q$, see equation (\ref{gen}), and $H$ is given
in (\ref{hf}) here. After some algebra, one can relate all the
ZZ quantities to the present ones. For example,
\[
c_1 = \frac{\lambda_p}{2 n} \; (4 (D - 2) a - c_3) + 2 q 
\; , \; \; \; c_3 = 4 (a - a_0) \; .
\]
These relations can be inverted to obtain $(a_0, a, q)$ in terms
of $(c_1, c_3, k_{ZZ})$. Using these inverse relations and after
a long algebra, one can show that the constraints (\ref{acq})
here, with $c$ eliminated from them, imply that
\[
k_{ZZ}^2 = \frac{(n + 1) \Delta}{n (D - 2)} - \frac{c_1^2 
\Delta}{2 (D - 2)} - \frac{(n + p) \Delta c_3^2}{4 
(D - 2)^2} + \frac{1}{4} \left( \lambda_p c_1 
+ \frac{n c_3}{D - 2} \right)^2 
\]
where $\Delta$ is as given in equation (\ref{delta}) here. It
can be checked that the above expression agrees with that given
in \cite{bmo} for $D = 10$, $\lambda_p = \frac{3 - p}{2}$, and
with that given in \cite{kam} for $c_3 = 0$.

%\vspace{3ex}

%{\bf Acknowledgement:} 
%Thank you. 

%\newpage


\begin{thebibliography}{999}

%\cite{Zhou:1999nm}
\bibitem{zz}
B.~Zhou and C.~J.~Zhu,
%``The complete black brane solutions in D-dimensional coupled gravity
%system,''
arXiv: hep-th/9905146.
%%CITATION = HEP-TH 9905146;%%

%\cite{Horowitz:1991cd}
\bibitem{solns}
G.~T.~Horowitz and A.~Strominger,
%``Black strings and P-branes,''
Nucl.\ Phys.\ B {\bf 360} (1991) 197; 
%%CITATION = NUPHA,B360,197;%%
\\
%\cite{Gueven:1992hh}
R.~Gueven,
%``Black p-brane solutions of D = 11 supergravity theory,''
Phys.\ Lett.\ B {\bf 276} (1992) 49;
%%CITATION = PHLTA,B276,49;%%
\\
%\cite{Lu:1993vt}
%\bibitem{adm}
J.~X.~Lu,
%``ADM masses for black strings and p-branes,''
Phys.\ Lett.\ B {\bf 313} (1993) 29
[arXiv: hep-th/9304159];
%%CITATION = HEP-TH 9304159;%%
\\
%\cite{Duff:1993ye}
M.~J.~Duff and J.~X.~Lu,
%``Black and super p-branes in diverse dimensions,''
Nucl.\ Phys.\ B {\bf 416} (1994) 301
[arXiv: hep-th/9306052];
%%CITATION = HEP-TH 9306052;%%
\\
%\cite{Duff:1994an}
M.~J.~Duff, R.~R.~Khuri and J.~X.~Lu,
%``String solitons,''
Phys.\ Rept.\  {\bf 259} (1995) 213
[arXiv: hep-th/9412184];
%%CITATION = HEP-TH 9412184;%%
\\
%\cite{Lu:1995hm}
H.~Lu, C.~N.~Pope, E.~Sezgin and K.~S.~Stelle,
%``Dilatonic p-brane solitons,''
Phys.\ Lett.\ B {\bf 371} (1996) 46
[arXiv: hep-th/9511203];
%%CITATION = HEP-TH 9511203;%%
\\
%\cite{Lu:1995yn}
H.~Lu and C.~N.~Pope,
%``p-brane Solitons in Maximal Supergravities,''
Nucl.\ Phys.\ B {\bf 465} (1996) 127
[arXiv: hep-th/9512012]; 
%%CITATION = HEP-TH 9512012;%%
\\
%\cite{Duff:1996hp}
M.~J.~Duff, H.~Lu and C.~N.~Pope,
%``The black branes of M-theory,''
Phys.\ Lett.\ B {\bf 382} (1996) 73
[arXiv: hep-th/9604052];
%%CITATION = HEP-TH 9604052;%%
\\
%\cite{Ohta:1997gw}
%\bibitem{Ohta:1997gw}
  N.~Ohta,
  %``Intersection rules for non-extreme p-branes,''
  Phys.\ Lett.\ B {\bf 403} (1997) 218
  [arXiv: hep-th/9702164]; 
  %%CITATION = HEP-TH 9702164;%%
\\
%\cite{Gal'tsov:1998yu}
D.~V.~Gal'tsov and O.~A.~Rytchkov,
%``Generating branes via sigma-models,''
Phys.\ Rev.\ D {\bf 58} (1998) 122001
[arXiv: hep-th/9801160];
%%CITATION = HEP-TH 9801160;%%
\\
%\cite{Stelle:1998xg}
K.~S.~Stelle,
%``BPS branes in supergravity,''
[arXiv: hep-th/9803116];
%%CITATION = HEP-TH 9803116;%%
\\
V. D. Ivashchuk and V. N. Melnikov, 
Class. Quantum Grav. {\bf 18} (2001) R87. 

%\cite{Brax:2000cf}
\bibitem{bmo}
P.~Brax, G.~Mandal and Y.~Oz,
%``Supergravity description of non-BPS branes,''
Phys.\ Rev.\ D {\bf 63} (2001) 064008
[arXiv: hep-th/0005242].
%%CITATION = HEP-TH 0005242;%%

%\cite{Kobayashi:2004ay}
\bibitem{kam}
S.~Kobayashi, T.~Asakawa and S.~Matsuura,
%``Open string tachyon in supergravity solution,''
arXiv: hep-th/0409044.
%%CITATION = HEP-TH 0409044;%%

%\cite{Bertolini:2000jy}
\bibitem{uses}
M.~Bertolini et al, 
%, P.~Di Vecchia, M.~Frau, A.~Lerda, R.~Marotta and R.~Russo,
%``Is a classical description of stable non-BPS D-branes possible?,''
Nucl.\ Phys.\ B {\bf 590} (2000) 471
[arXiv: hep-th/0007097]. 
%%CITATION = HEP-TH 0007097;%%

%\cite{Gal'tsov:2004kn}
\bibitem{glc}
D.~V.~Gal'tsov, J.~P.~S.~Lemos and G.~Clement,
%``Supergravity p-branes revisited: Extra parameters, uniqueness, and
%topological censorship,''
Phys.\ Rev.\ D {\bf 70} (2004) 024011
[arXiv: hep-th/0403112].
%%CITATION = HEP-TH 0403112;%%

%\cite{Lu:2004dp}
\bibitem{lr}
J.~X.~Lu and S.~Roy,
%``Supergravity approach to tachyon condensation on the brane-antibrane
%system,''
Phys.\ Lett.\ B {\bf 599} (2004) 313
[arXiv: hep-th/0403147];
%%CITATION = HEP-TH 0403147;%%
\\
%\cite{Lu:2004ms}
%\bibitem{Lu:2004ms}
%J.~X.~Lu and S.~Roy,
%``Static, non-SUSY p-branes in diverse dimensions,''
JHEP {\bf 02} (2005) 001
arXiv: hep-th/0408242; 
%%CITATION = HEP-TH 0408242;%%
\\
%\cite{Lu:2004xi}
%\bibitem{Lu:2004xi}
%J.~X.~Lu and S.~Roy,
%``Delocalized, non-SUSY p-branes, tachyon condensation and tachyon matter,''
JHEP {\bf 11} (2004) 008
[arXiv: hep-th/0409019];
%%CITATION = HEP-TH 0409019;%%
\\
%\cite{Lu:2005ju}
%\bibitem{Lu:2005ju}
%J.~X.~Lu and S.~Roy,
%``Non-SUSY $p$-branes delocalized in two directions, 
% tachyon condensation and T-duality,''
{[}arXiv: hep-th/0503007].
%%CITATION = HEP-TH 0503007;%%

\bibitem{mo}
Y.~G.~Miao and N.~Ohta,
%``Complete intersecting non-extreme p-branes,''
Phys.\ Lett.\ B {\bf 594} (2004) 218 
[arXiv: hep-th/0404082]. 
%%CITATION = HEP-TH 0404082;%%
\\
See also 
%\cite{Edelstein:2004tp}
%\bibitem{em}
J.~D.~Edelstein and J.~Mas,
%``Localized intersections of non-extremal p-branes and S-branes,''
JHEP {\bf 06} (2004) 015
[arXiv: hep-th/0403179].
%%CITATION = HEP-TH 0403179;%%

%\cite{Tseytlin:1996bh}
\bibitem{rules}
A.~A.~Tseytlin,
%``Harmonic superpositions of M-branes,''
Nucl.\ Phys.\ B {\bf 475} (1996) 149
[arXiv: hep-th/9604035]; 
%%CITATION = HEP-TH 9604035;%%
\\
%\cite{Cvetic:1996gq}
M.~Cvetic and A.~A.~Tseytlin,
%``Non-extreme black holes from non-extreme intersecting M-branes,''
Nucl.\ Phys.\ B {\bf 478} (1996) 181
[arXiv: hep-th/9606033];
%%CITATION = HEP-TH 9606033;%%

%\cite{Russo:1996if}
\bibitem{boost}
J.~G.~Russo and A.~A.~Tseytlin,
%``Waves, boosted branes and BPS states in M-theory,''
Nucl.\ Phys.\ B {\bf 490} (1997) 121
[arXiv: hep-th/9611047];
%%CITATION = HEP-TH 9611047;%%
\\
%\cite{Tseytlin:1996zb}
A.~A.~Tseytlin,
%``On the structure of composite black p-brane configurations and related
%black holes,''
Phys.\ Lett.\ B {\bf 395} (1997) 24
[arXiv: hep-th/9611111];
%%CITATION = HEP-TH 9611111;%%
\\
%\cite{Ohta:1997wp}
%\bibitem{Ohta:1997wp}
  N.~Ohta and T.~Shimizu,
  %``Non-extreme black holes from intersecting M-branes,''
  Int.\ J.\ Mod.\ Phys.\ A {\bf 13} (1998) 1305
  [arXiv: hep-th/9701095]; 
  %%CITATION = HEP-TH 9701095;%%
\\
%\cite{Ohta:1997wd}
%\bibitem{Ohta:1997wd}
  N.~Ohta and J.~G.~Zhou,
  %``Towards the classification of non-marginal bound states of M-branes and
  %their construction rules,''
  Int.\ J.\ Mod.\ Phys.\ A {\bf 13} (1998) 2013
  [arXiv: hep-th/9706153];
  %%CITATION = HEP-TH 9706153;%%
\\
%\cite{Das:1997tk}
S.~R.~Das, S.~D.~Mathur, S.~Kalyana Rama and P.~Ramadevi,
%``Boosts, Schwarzschild black holes and absorption cross-sections in M
%theory,''
Nucl.\ Phys.\ B {\bf 527} (1998) 187
[arXiv: hep-th/9711003]. 
%%CITATION = HEP-TH 9711003;%%

%\cite{Bergshoeff:1995as}
\bibitem{bho}
E.~Bergshoeff, C.~M.~Hull and T.~Ortin,
%``Duality in the type II superstring effective action,''
Nucl.\ Phys.\ B {\bf 451} (1995) 547
[arXiv: hep-th/9504081].
%%CITATION = HEP-TH 9504081;%%

%\cite{Bagchi:2004fe}
\bibitem{bk}
A.~Bagchi and S.~Kalyana Rama,
%``Cosmology and static spherically symmetric solutions 
%in D-dimensional scalar
%tensor theories: Some Novel features,''
Phys.\ Rev.\ D {\bf 70} (2004) 104030
[arXiv: gr-qc/0408030];
%%CITATION = GR-QC 0408030;%%
\\
%\cite{KalyanaRama:1995zu}
%\bibitem{KalyanaRama:1995zu}
S.~Kalyana Rama and S.~Ghosh,
%``Short distance repulsive gravity as a consequence of nontrivial PPN
%parameters Beta and gamma,''
Phys.\ Lett.\ B {\bf 383} (1996) 31
%[Phys.\ Lett.\ B {\bf 384} (1996) 50]
[arXiv:hep-th/9505167].
%%CITATION = HEP-TH 9505167;%%

%\cite{Capozziello:1993tr}
\bibitem{cosm}
S.~Capozziello and R.~de Ritis,
%``Scale factor duality and general transformations for string cosmology,''
Int.\ J.\ Mod.\ Phys.\ D {\bf 2} (1993) 367; 
%%CITATION = IMPAE,D2,367;%%
\\
%\cite{Lidsey:1995ft}
J.~E.~Lidsey,
%``Scale Factor Duality and Hidden Supersymmetry in Scalar--Tensor Cosmology,''
Phys.\ Rev.\ D {\bf 52} (1995) 5407
[arXiv: gr-qc/9510017]; 
%%CITATION = GR-QC 9510017;%%
\\
%\cite{Clancy:1998ka}
D.~Clancy, J.~E.~Lidsey and R.~K.~Tavakol,
%``Scale factor dualities in anisotropic cosmologies,''
Class.\ Quant.\ Grav.\  {\bf 15} (1998) 257
[arXiv: gr-qc/9802053];
%%CITATION = GR-QC 9802053;%%
\\
%\cite{Gasperini:2002bn}
M.~Gasperini and G.~Veneziano,
%``The pre-big bang scenario in string cosmology,''
Phys.\ Rept.\  {\bf 373} (2003) 1
[arXiv: hep-th/0207130].
%%CITATION = HEP-TH 0207130;%%

\bibitem{sbrane}
%\cite{Gutperle:2002ai}
M.~Gutperle and A.~Strominger,
%``Spacelike branes,''
JHEP {\bf 04} (2002) 018
[arXiv: hep-th/0202210]; 
%%CITATION = HEP-TH 0202210;%%
\\
%\cite{Chen:2002yq}
C.~M.~Chen, D.~V.~Gal'tsov and M.~Gutperle,
%``S-brane solutions in supergravity theories,''
Phys.\ Rev.\ D {\bf 66} (2002) 024043
[arXiv: hep-th/0204071];
%%CITATION = HEP-TH 0204071;%%
\\
%\cite{Kruczenski:2002ap}
M.~Kruczenski, R.~C.~Myers and A.~W.~Peet,
%``Supergravity S-branes,''
JHEP {\bf 05} (2002) 039
[arXiv: hep-th/0204144];
%%CITATION = HEP-TH 0204144;%%
\\
%\cite{Roy:2002ik}
S.~Roy,
%``On supergravity solutions of space-like Dp-branes,''
JHEP {\bf 08} (2002) 025
[arXiv: hep-th/0205198];
%%CITATION = HEP-TH 0205198;%%
\\
%\cite{Ohta:2003uw}
%\bibitem{Ohta:2003uw}
  N.~Ohta,
  %``Intersection rules for S-branes,''
  Phys.\ Lett.\ B {\bf 558} (2003) 213
  [arXiv: hep-th/0301095].
  %%CITATION = HEP-TH 0301095;%%


\end{thebibliography}
\end{document}